%% file: wc-paper-final.tex
\documentclass[letterpaper,twocolumn,10pt]{article}

\usepackage{usenix-2020-09} %

\input{preamble.tex}

\begin{document}

\input{meta.tex}

\author{
    {\rm Andrew Jeffery}\\
    University of Cambridge
    \and
    {\rm Chris Jensen}\\
    University of Cambridge
    \and
    {\rm Richard Mortier}\\
    University of Cambridge
} %

\input{index.tex}

\end{document}

%% file: preamble.tex
\usepackage{booktabs}

\usepackage{csquotes}

\usepackage{amsmath}

\usepackage{subcaption}

\usepackage{tcolorbox}

\usepackage{algpseudocode}
\usepackage{algorithm}

\usepackage{xurl}

\def\vCPU/{{vCPU}}
\def\CPU/{{CPU}}
\def\cgroup/{{\verb|cgroup|}}

\usepackage{xcolor}

%% file: meta.tex
\date{}

\title{\Large \bf Reducing Tail Latencies Through Environment- \\ and Neighbour-aware Thread Management}

%% file: index.tex
\maketitle

\begin{abstract}
    \input{abstract}
\end{abstract}

\input{introduction}

\input{system-model}

\input{overcommitment}

\input{autotuning}

\input{burst}

\input{neighbours}

\input{related}

\input{conclusion}

\bibliographystyle{plain}
\bibliography{paper}

%% file: abstract.tex
Application tail latency is a key metric for many services, with high latencies being linked directly to loss of revenue.
Modern deeply-nested micro-service architectures exacerbate tail latencies, increasing the likelihood of users experiencing them.
In this work, we show how CPU overcommitment by OS threads leads to high tail latencies when applications are under heavy load.
CPU overcommitment can arise from two operational factors: incorrectly determining the number of CPUs available when under a CPU quota, and the ignorance of neighbour applications and their CPU usage.
We discuss different languages' solutions to obtaining the CPUs available, evaluating the impact, and discuss opportunities for a more unified language-independent interface to obtain the number of CPUs available.
We then evaluate the impact of neighbour usage on tail latency and introduce a new neighbour-aware threadpool, the friendlypool, that dynamically avoids overcommitment.
In our evaluation, the friendlypool reduces maximum worker latency by up to $6.7\times$ at the cost of decreasing throughput by up to $1.4\times$.

%% file: introduction.tex
\section{Introduction}

As shown repeatedly by large companies including Google, Amazon, and independent studies~\cite{makedatauseful, costoflatency, mayerweb2.0, autosens, hft}: increased latency reduces client retention and costs money.
This is exacerbated by modern micro-service architectures due to their deep nesting~\cite{crisp, microarch}, leading to a higher likelihood of a tail latency being observed by a user.
Bursts of demand may also be correlated~\cite{internetbursts}, leading to resource demand surges.

\begin{table}
    \centering
    \caption{Overview of the biggest VMs by CPU core count available at large cloud providers.}\label{tab:manycorevms}
    \begin{tabular}{lrl}
        \toprule
        Cloud provider & vCPUs & Instance type           \\
        \midrule
        AWS            & 192   & \verb|c7i.metal-48xl|   \\
        Azure          & 96    & \verb|Standard_D96d_v5| \\
        GCP            & 360   & \verb|c3d-standard-360| \\
        \bottomrule
    \end{tabular}
\end{table}

As Table~\ref{tab:manycorevms} shows, large CPU counts are now available at many cloud providers.
This leads to applications that can scale to the number of CPUs available on the host in order to handle larger volumes of requests~\cite{cassandrascale, hadoopscale, sparkle}.
However, they assume that they operate alone on the host, and so can make full use of the CPU resources.
This assumption is often false, leading to overcommitment~\cite{mindthegap}.
Each application creates the number of CPUs worth of threads, meaning the total threads per CPU scales with the number of applications.
This leads to contention over the CPUs, leaving each application with far less than the full host it was expecting to have.

CPU quotas are used to isolate applications on the same host from using too much CPU time~\cite{cpuquota}.
However, the interfaces that many applications use for determining the available CPU resources do not encapsulate the CPU quota's restrictions~\cite{schedgetaffinity, nproc}.
This is in part due to the isolation only being lightweight and so a leaky abstraction~\cite{cgroups, namespaces}.
This directly leads to CPU overcommitment and high tail latencies.

CPU quotas also prevent an application from bursting its CPU usage in order to handle demand surges.
This leads to requests being queued, increasing the latency observed, or rejecting requests outright.

To take advantage of the larger nodes' size and cost-efficiency applications are packed tightly, aiming for high utilisation~\cite{borg}.
However, when the these applications have coinciding demand bursts they can contend over shared CPUs.
Since the number of CPUs available to an application is typically treated as a static constant they do not adapt to the reduced CPU usage that they obtain.
The CPU overcommitment then leads to increased tail latencies.

In this work we show that CPU overcommitment is still a problem in modern contexts (Section~\ref{sec:background}).
We then draw comparisons and corresponding evaluations from this for the contributions of CPU quotas (Section~\ref{sec:autotune} and~\ref{sec:bursting}) and neighbour CPU load (Section~\ref{sec:neighbours}).
Finally, we present a new thread pool design that dynamically avoids overcommitment, improving tail latencies (Section~\ref{sec:friendlypool}).

%% file: system-model.tex
\section{Overcomitment: OS Threads > CPUs}\label{sec:background}

This section outlines the model of computation that we refer to throughout this work, as well as defining overcommitment and evaluating its impacts on tail latency.

\paragraph{Computation Model}

A single processor socket is formed of multiple physical processing cores, each presenting potentially multiple CPUs due to hyper-threading~\cite{lscpu}.
Each CPU is exposed to the operating system (OS), which manages the execution of OS threads over the many CPUs.
These OS threads are \emph{scheduled} onto the CPUs based on a scheduling algorithm.
OS threads are preemptive, so they may not complete their computation in one scheduling cycle before being taken off the CPU.\@

\verb|cgroups|~\cite{cgroups} enable the assignment of CPU shares and a CPU quota to a process.
The OS scheduler, Linux's CFS in this case, packs OS threads to CPUs based on their share relative to all total shares of OS threads ready to be executed.
The scheduler also enforces the quota as the maximum amount of time a process is able to run for in any given period.
CPU quotas are typically employed to prevent processes from hogging CPU resources, particularly in multi-tenant environments.

Since OS threads are typically heavy-weight, modern languages use a design of \emph{lightweight threads}~\cite{lightweightthreads}.
These lightweight threads are language-specific so require a runtime to manage them, multiplexing them over OS threads created by the runtime.
The runtime's OS threads are then commonly partitioned into \emph{workers} and \emph{waiters}.
Worker threads are small in number and perform the CPU-bound work, while waiter threads are greater in number and execute blocking operations, spending most of their time in wait queues.

%% file: overcommitment.tex
\paragraph{Overcommitment}

A well known result is that as the number of threads increases, coordination costs also increase, due to contention over shared resources~\cite{multicorelocks}.
This also extends to CPUs as once each \CPU/ is fully utilized, adding further threads increases coordination costs without increasing the amount of useful work done.
This situation is known as overcommitment.
However, considering the total quantity of useful work is not the whole story, instead focusing here on latency.

\begin{figure}
    \centering
    \includegraphics[width=0.6\linewidth]{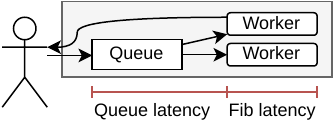}
    \caption{Structure of the applications and the latencies being measured.}\label{fig:structure-app}
\end{figure}

\begin{figure}
    \begin{subfigure}{\columnwidth}
        \includegraphics[width=\linewidth]{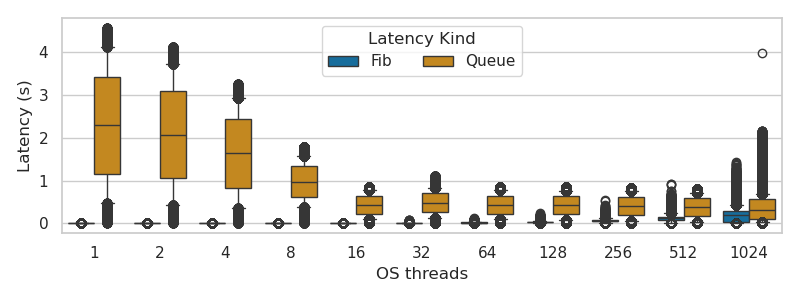}
        \caption{Queue latency decreases as more threads are used, up to the number of CPUs. Fib latency remains low until overcommitment.}
    \end{subfigure}

    \begin{subfigure}{\columnwidth}
        \includegraphics[width=\linewidth]{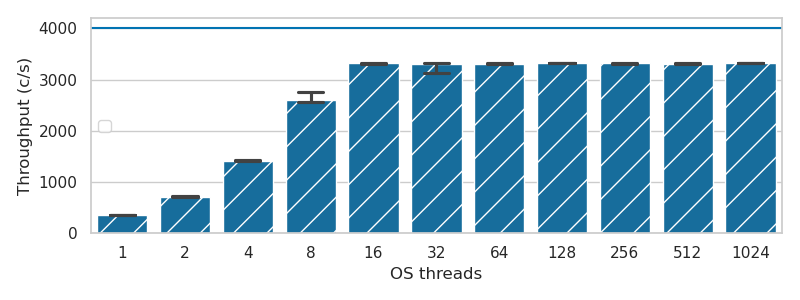}
        \caption{Throughput gradually increases as more threads are used, up to the number of CPUs.}
    \end{subfigure}

    \caption{Overall latency and throughput at various amounts of OS threads in Rust. Workers have no contention.}\label{fig:overcommit-nocontention}
\end{figure}

\begin{figure}
    \begin{subfigure}{\columnwidth}
        \includegraphics[width=\linewidth]{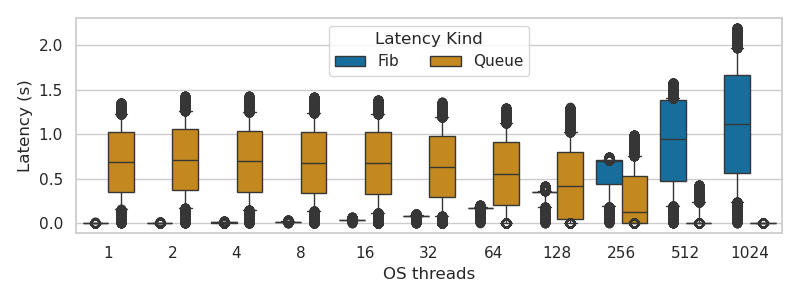}
        \caption{Queue latency decreases as more threads are used. Fib latency increases with overcommitment.}
    \end{subfigure}

    \begin{subfigure}{\columnwidth}
        \includegraphics[width=\linewidth]{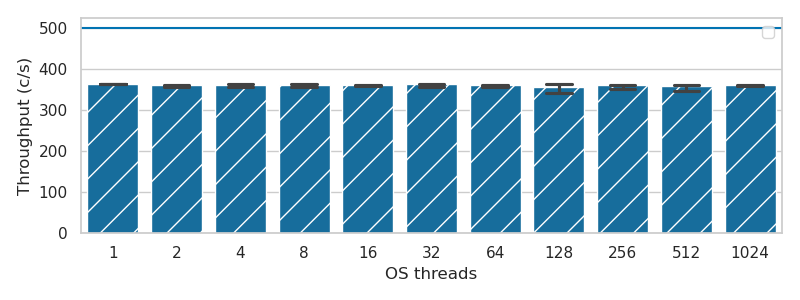}
        \caption{Throughput remains constant, limited by lock contention.}\label{fig:overcommit-throughput-contention}
    \end{subfigure}

    \caption{Overall latency and throughput at various amounts of OS thread overcommitment in Rust. Workers have contention over the \texttt{fib} computation starting with a lock at \texttt{fib(30)}.}\label{fig:overcommit-contention} %
\end{figure}

To study the effect of overcommitment throughout this work we implement a simple program that spawns worker threads, each performing CPU-bound work, a naive fibonacci (\verb|fib|) sequence computation, computing \verb|fib(30)|.
We drive the workers at a target rate in an open-loop fashion, representing a uniform demand.
We collect timing measurements at three points, the time at which the work item is \enquote{received} by the service and pushed into the queue (\emph{queue start}), the time the work item is pulled out of the queue by a worker (\emph{fib start}), and the time that the work item has finished being processed (\emph{end}).
From these we derive the \emph{queue latency} and the \emph{fib latency}, shown in Figure~\ref{fig:structure-app}.
We also calculate overall throughput of the process as the time taken to complete divided by the total runtime.

The experiments in this paper were conducted on a single machine running Linux 6.2.0 with a dual socket Intel Xeon Silver 4112 Processor featuring 4 physical cores each with hyper-threading, so 16 CPUs total.
Experiments are run for 5 seconds with 5 repeats.
Latency plots show points for all repeats, throughput plots show the median with an error bar between the minimum and maximum across repeats.

Figure~\ref{fig:overcommit-nocontention} breaks down the total throughput and latency for client requests at various numbers of threads in a Rust implementation.
When the total system's throughput is lower than the request rate (fewer than 16 threads) requests must queue, resulting in higher than optimal latency.
However if the system is overcommitted (more than 16 threads), each thread has a lower effective throughput and hence increased fib latency, getting worse with more overcommitment.
This occurs because the worker thread pulls an item off the queue, starting its fib latency timer, but due to overcommitment, the OS scheduler may choose to preempt the worker thread and run another before the current one completes working on its item.
The original thread must then wait to be scheduled onto a CPU again to continue work and report its result, stopping the timer.
The effect of overcommitment on the latency of each request is exacerbated by contention over shared resources, a mutex around the computation of \verb|fib(30)| in this case, shown in Figure~\ref{fig:overcommit-contention}.
Since this is an extreme contention scenario, showing the limit, we proceed only without contention to study the most optimistic case.

In the rest of this work we will not consider queue latency as it arises due to the overload of the service and other strategies such as load shedding or horizontal scaling can be used to mitigate it.

%% file: autotuning.tex
\section{How Do Applications Tune Themselves?}\label{sec:autotune}

When an application starts up it needs to perform some initialisation.
This typically involves getting some information about the environment and using that to set up some internal configuration parameters, such as threadpool sizes.
This section investigates how this initialisation works under the presence of CPU quotas and the impacts that it can have.

\subsection{Working Under CPU Quotas}

\begin{tcolorbox}
    \textbf{Assuming:} Quotas in place.

    \textbf{Problem 1:} Applications do not account for their CPU quota when calculating the number of CPUs.
\end{tcolorbox}

The logical conclusion from the previous section is that applications should limit themselves to OS threads = \CPU/s, a thread-per-core architecture~\cite{threadpercore}, to avoid overcommitment which should be easy to read from the environment.

\begin{table}
    \centering
    \caption{Lightweight threading mechanism, runtime, and CPU source for each language.}\label{tab:cpu-info-mechanism}
    \begin{tabular}{llll}
        \toprule
        Lang  & Thread                            & Runtime            & CPU source   \\
        \midrule
        C++   & Coroutine~\cite{coroutine}        & Built-in           & CPU affinity \\
        Go    & Goroutine~\cite{goroutine}        & Built-in           & CPU affinity \\
        Java  & Virt.\ threads~\cite{javathreads} & Built-in           & cgroup       \\
        Julia & Task~\cite{juliatasks}            & Built-in           & CPU affinity \\
        OCaml & Eio fiber~\cite{eiofiber}         & Eio~\cite{eio}     & CPU affinity \\
        Rust  & Tokio task~\cite{tokiotask}       & Tokio~\cite{tokio} & cgroup       \\
        \bottomrule
    \end{tabular}
\end{table}

\begin{figure}
    \centering
    \begin{subfigure}{0.45\columnwidth}
        \includegraphics[width=\linewidth]{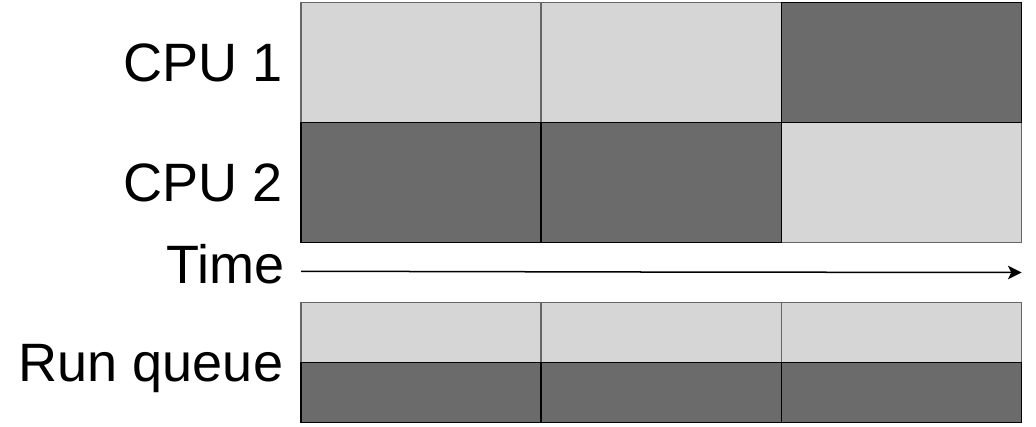}
        \caption{CPU quota, shared.}\label{fig:quota-shared}
    \end{subfigure}
    \begin{subfigure}{0.45\columnwidth}
        \includegraphics[width=\linewidth]{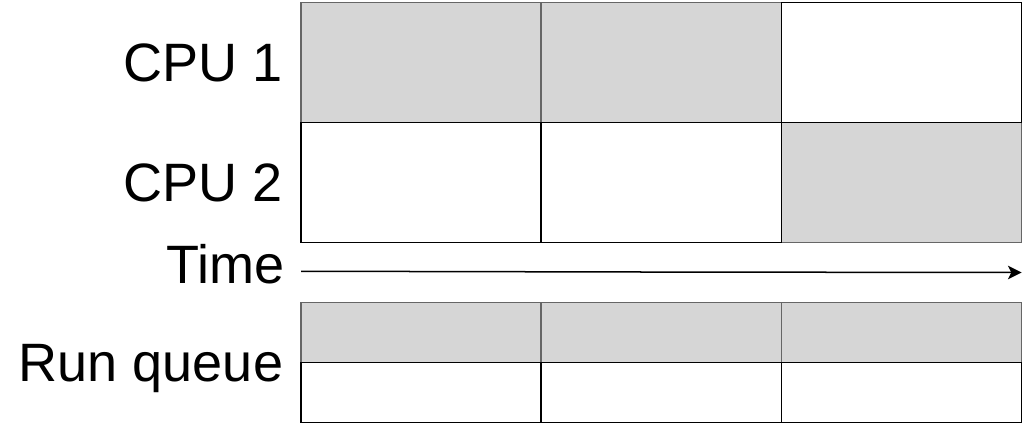}
        \caption{CPU quota, alone.}\label{fig:quota-alone}
    \end{subfigure}
    \caption{Example schedulings, shown in scheduling periods of 2 apps with CPU quotas equivalent to 1 CPU on a 2 CPU system.}
\end{figure}

Unfortunately, when applications are deployed with CPU quotas, obtaining the correct CPU count becomes more difficult.
We would expect that an application should use the formula: OS threads = quota cores, for its self-tuning, where quota cores is \(\text{quota time} / \text{scheduler period}\).
However, despite the seeming ease of obtaining this from the \verb|cgroups| API, multiple languages do not obtain the correct number.
Table~\ref{tab:cpu-info-mechanism} summarizes the approaches some popular languages use to obtain the number of CPUs available for the application.
We observed that a large number of languages are not \verb|cgroup| aware, and so do not look for their allocated CPU quota.
Instead, most rely on the OS system calls, such as \verb|sched_getaffinity|~\cite{schedgetaffinity} and \verb|sysconf(_SC_NPROCESSORS_ONLN)|~\cite{nproc}, which do not reflect the CPU quota in place, instead returning the total number of CPUs in the system.

\begin{figure}
    \begin{subfigure}{\linewidth}
        \includegraphics[width=\linewidth]{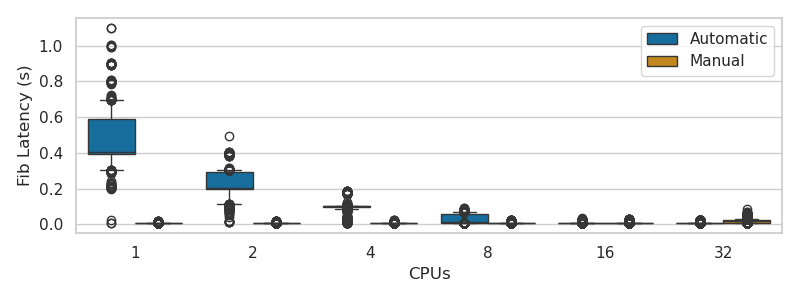}
        \caption{Fib latency remains constantly low when matching threads to CPU quota. Letting the runtime decide the number of threads leads to a large fib latency, decreasing as CPU quota approaches number of CPUs.}\label{fig:quota-latency}
    \end{subfigure}

    \begin{subfigure}{\linewidth}
        \includegraphics[width=\linewidth]{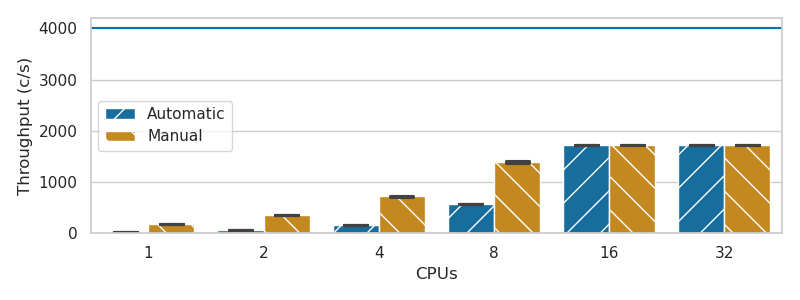}
        \caption{Throughput is impacted by overcommitment of CPUs due to incorrect automatic detection of CPU availability.}\label{fig:quota-throughput}
    \end{subfigure}

    \caption{Impact of using an incorrect OS thread count when under a CPU quota in Go.}\label{fig:quota-perf}
\end{figure}

Since the CPU quota is an enforcement on maximum usage, overcommitment of OS threads can easily happen when incorrectly scaled, as shown in Figure~\ref{fig:quota-shared} by the full run queue.
As shown in Section~\ref{sec:background}, overcommitment of OS threads to CPUs results in increased latency due to preemption.
Figure~\ref{fig:quota-perf} presents the fib latency and the throughput from Go when manually setting the optimal number of threads and when letting the runtime determine it automatically.
However, in this experiment we place the applications in \verb|cgroups| with CPU quotas corresponding to varying numbers of CPUs.
Since Go, the cloud-native language~\cite{gocloudnative}, is not \verb|cgroup|-aware it spawns a number of OS threads equal to the number of CPUs on the entire machine (16 in this case).
This leads Go to having a large overcommitment for most cases and sees an accordingly high latency on its work.
For the Go language, this has been observed widely, including at Uber, a prominent user of Go, having published their library to obtain the CPU quota from the \verb|cgroup| API to contain the issue~\cite{automaxprocs}.

Notably, other CPU containment mechanisms do not feature these limitations.
CPU shares, which govern a weighting for the time share a thread gets, are only a lower bound.
CPU affinity allows pinning a thread to a CPU but, working on the thread-level, does not introduce a direct limitation on how many threads can run.

\subsection{Getting the Correct CPU Count}

\begin{tcolorbox}
    \textbf{Assuming:} Quotas in place.

    \textbf{Solution 1:} Fix the syscall for number of CPUs to be quota-aware, or add a new one.
\end{tcolorbox}

Given that we know the interfaces languages use to determine the number of CPUs to use we see two complementary solutions.
Firstly, languages and their runtimes could be made \verb|cgroup|-aware.
This is likely the fastest method to fix this situation but requires that each language independently work out their fix and correctly interpret the \verb|cgroup| API.\@
It also places maintenance burden on the developers as new versions of the \verb|cgroup| API are added, such as with \verb|cgroups| v1 and v2.
Secondly, as most languages already see fit to offload the calculation to the OS, a new \emph{syscall} could be added to obtain the correct number of CPUs that the process should be using.
This would need to take into account the CPU affinity mask for the process as well as implications of a \verb|cgroups| CPU quota.
This solution would be portable between languages and reduce the complexity required in them.
Importantly, this would encapsulate being \verb|cgroup|-aware but also aim to incorporate future changes too.
We leave the implementation of these fixes out of the scope of this work, but note that the results of their changes would follow the pattern shown by the \enquote{Manual} thread count in Figure~\ref{fig:quota-perf}.

%% file: burst.tex
\section{Capitalising On Spare Resources}\label{sec:bursting}

Assuming that the issue with CPU quotas has been fixed, we now look at limitations that are still imposed.
To do this, we extend our view broader than our single application, out to the host and all the processes it may be running.

\subsection{CPU Quotas Are Wasteful}

\begin{tcolorbox}
    \textbf{Assuming:} Quotas in place, OS threads = quota cores.

    \textbf{Problem 2:} Applications without sufficiently busy neighbours can see wasted CPUs that they can't use.
\end{tcolorbox}

When all applications are fully using their quota and the sum of the quota cores is equal to the number of CPUs then the system behaves well, each OS thread gets matched to a CPU.\@
However, such a case with constant work is unlikely, even for batch jobs.
When an application does not fully utilise its quota then some portion of the CPU will be left under-utilised, shown in Figure~\ref{fig:quota-alone}.
This wastage is clearly suboptimal from a utilisation perspective and would ideally be avoided.
Particularly this involves the application being burstable, often in response to its demand.
This is more common in modern deployments with dense packing of applications to hosts.

\subsection{Enabling Bursts}

\begin{tcolorbox}
    \textbf{Assuming:} Quotas in place, OS threads = quota cores.

    \textbf{Solution 2:} Do not assign CPU quotas.
\end{tcolorbox}

In order to solve the issue of applications with quotas being unable to burst up their usage we simply recommend removing the CPU quota unless strictly required.
They have been shown to have negative performance implications in real-world contexts~\cite{avoidthrottling}.
This can also be used as a stopgap solution until better support for detecting the number of CPUs based off CPU quotas is added (Section~\ref{sec:autotune}).

%% file: neighbours.tex
\section{The Cost Of Ignoring Your Neighbours}\label{sec:neighbours}

This section analyses issues arising when other applications (neighbours) are also present on the host.
We assume that CPU quotas are not in place for the applications, but the scheduler is fair, as is Linux's CFS.\@

\begin{figure}
    \centering
    \begin{subfigure}{0.45\columnwidth}
        \includegraphics[width=\linewidth]{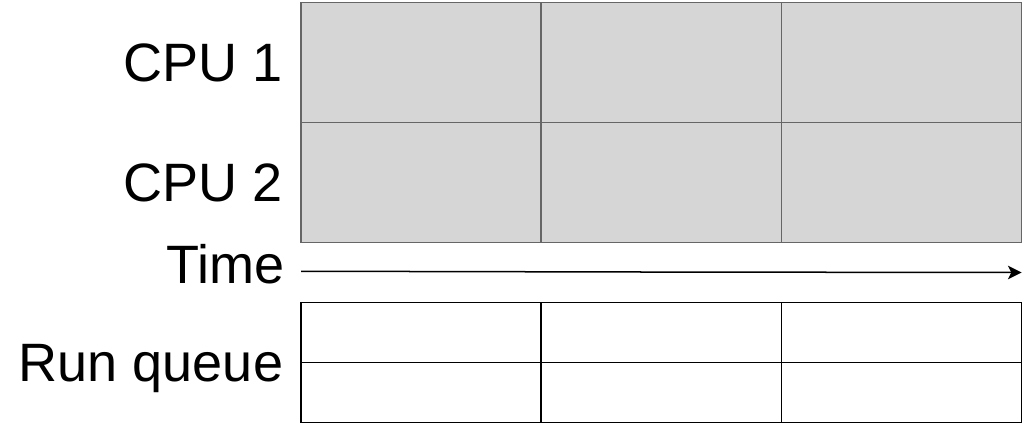}
        \caption{No quota, alone.}\label{fig:noquota-alone}
    \end{subfigure}
    \begin{subfigure}{0.45\columnwidth}
        \includegraphics[width=\linewidth]{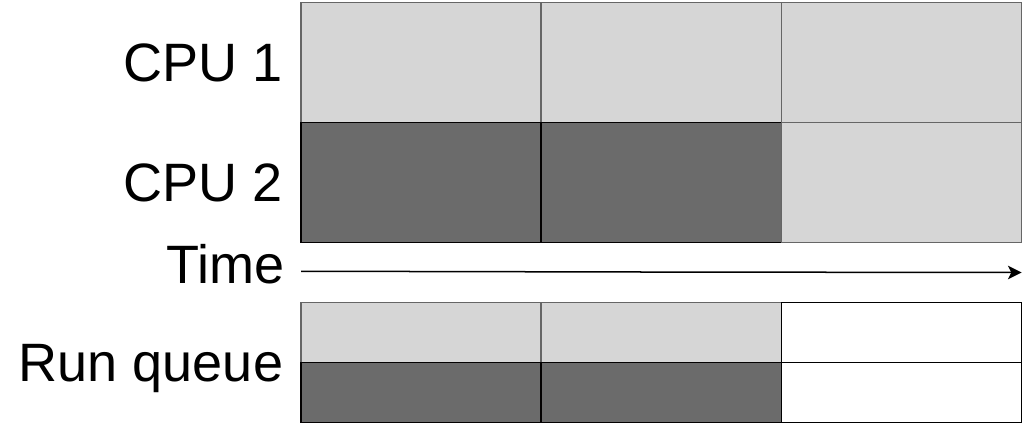}
        \caption{No quota, shared.}\label{fig:noquota-shared}
    \end{subfigure}
    \caption{Example schedulings, shown in scheduling periods of 2 apps with no CPU quotas on a 2 CPU system.}
\end{figure}

\subsection{Noisy Neighbourhood}

\begin{tcolorbox}
    \textbf{Assuming:} No quotas in place, OS Threads = CPUs.

    \textbf{Problem 3:} Applications ignore their neighbours' usage when calculating the number of threads to use.
\end{tcolorbox}

Having removed the CPU quotas on applications, each application is free to create a number of threads equal to the number of CPUs as shown in Figure~\ref{fig:noquota-alone}.
Provided that the sum of busy threads at any instant in time does not exceed the number of CPUs the scheduler should be able to map each thread to its own distinct CPU.\@
However, in reality the overcommitment of OS threads to CPUs is severely higher than 1, by a factor of the number of applications, as each will spawn one thread per CPU, shown in Figure~\ref{fig:noquota-shared}.
Unknowingly to each application, there is contention over the CPU cores.

CPU shares should be used to ensure that each application, when it is busy, has a minimum portion of CPU time.
Assuming that all applications on a host are busy at the same time, then they will each be given their CPU shares.
In this scenario each application is then directly subject to its own overcommitment.
This effectively partitions the applications from each other, pushing the impact of overcommitment directly, and solely, onto them.
This means that applications that do not overcommit their share will not observe the associated increased latencies.
However, those applications that do overcommit will only impede the performance of their own other threads, and so observe the associated increase in latencies seen previously in Figures~\ref{fig:overcommit-nocontention} and~\ref{fig:overcommit-contention}.

\begin{figure}
    \begin{subfigure}{\linewidth}
        \includegraphics[width=\linewidth]{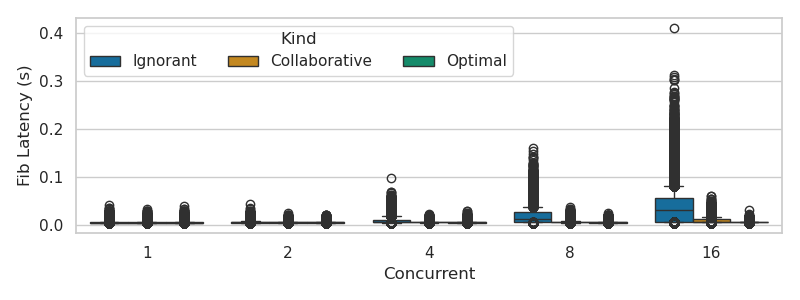}
        \caption{Ignoring neighbour usage leads to a large fib latency, increasing with number of neighbours. A collaborative approach reduces this.}\label{fig:concurrent-fib-latency}
    \end{subfigure}

    \begin{subfigure}{\linewidth}
        \includegraphics[width=\linewidth]{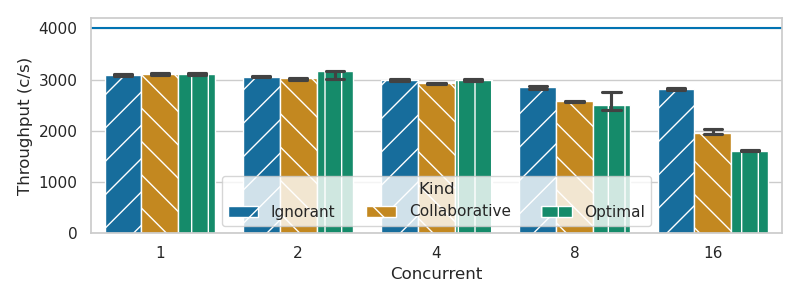}
        \caption{Ignoring neighbour usage maintains a high throughput, whereas partitioning CPUs by reducing thread count limits it.}\label{fig:concurrent-fib-throughput}
    \end{subfigure}

    \caption{Performance when sharing a node with neighbours using the Rust app. Ignorant: threads = CPUs for each process. Collaborative: the dynamic friendly threadpool. Optimal: threads = CPUs / processes. This highlights the trade-off space between latency and throughput from overcommitment.}\label{fig:concurrent-fib}
\end{figure}

There is now a conflict, between having enough OS threads available to be able to handle bursts when neighbours are not fully utilising their share, and not impacting latency when neighbours are busy.
Figure~\ref{fig:concurrent-fib} shows the effect on latency of overcommitting OS threads to CPUs when neighbours are busy, looking at the \enquote{Ignorant} kind.

\subsection{Dynamically Adapting To Neighbours}\label{sec:friendlypool}

\begin{tcolorbox}
    \textbf{Assuming:} No quotas in place, OS Threads = CPUs.

    \textbf{Solution 3:} Dynamically alter the number of OS threads that perform work for an application.
\end{tcolorbox}

To handle the adaptive nature of neighbour usage the pool of worker threads needs to dynamically scale up and down.
For this, we introduce the \emph{friendlypool}, a threadpool that is friendly to its neighbours and is aware of their usage, marked as \emph{Collaborative} in Figures.
It works by simply spawning another thread that is in charge of periodically obtaining the number of the CPUs that the current process should use and scaling the threadpool to that number.
When scaling, we do not destroy or create OS threads, but rather they each watch the number of desired active threads and pull work or not, accordingly.
Initially one OS thread is created per CPU.\@

The number of OS threads the current application should be using is equivalent to its proportion of the usage of the CPUs in the last period.
For instance, if the application is alone on the host, then it will be consuming all of the CPU, and so the usage will be 100\%.
However, say another identical application is run, then the usages may be evenly split, each 50\%.
This is a key property obtained from the fairness of the OS thread scheduler.
This usage can then be scaled by the number of CPUs in the system to count how many OS threads should be active.
The calculation is:

\[ \text{active threads} = \left\lceil O \cdot \frac{\text{cpu time}_{\text{self}}}{\text{cpu time}_{\text{all}}} \cdot \text{CPUs} \right\rceil \]

A \emph{control thread} polls the \enquote{cpu time} values periodically, limited by how often they are updated by the Kernel, the clock tick frequency.
This defaults to 100Hz in our setup, and so this is evaluated every 10ms.
When a worker thread has no work to perform, it \emph{parks} itself, waiting for the control thread to \emph{unpark} it when it should execute.

\begin{figure}
    \begin{subfigure}{\linewidth}
        \includegraphics[width=\linewidth]{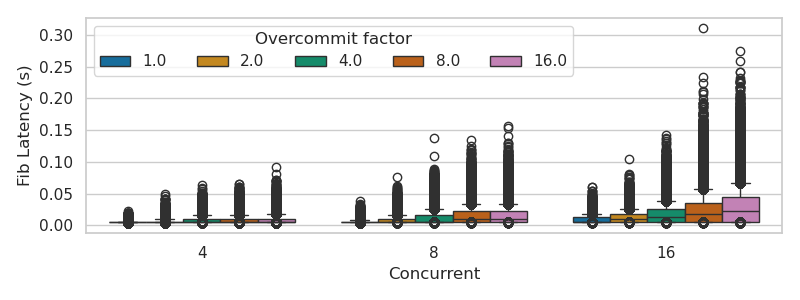}
        \caption{The overcommit factor is directly proportional to the fib latency.}
    \end{subfigure}

    \begin{subfigure}{\linewidth}
        \includegraphics[width=\linewidth]{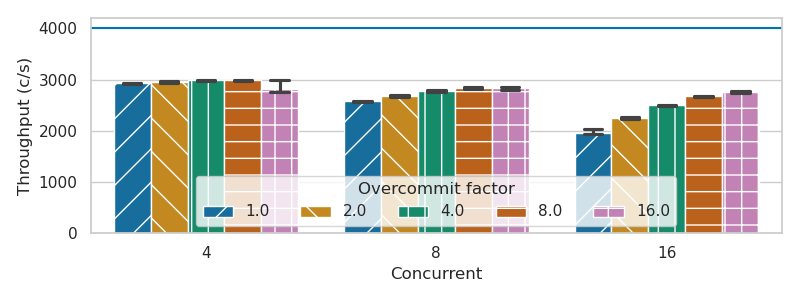}
        \caption{The overcommit factor is directly proportional to the throughput.}
    \end{subfigure}

    \caption{Performance of the collaborative friendly pool with variable overcommitment factor.}\label{fig:concurrent-overcommit}
\end{figure}

Figure~\ref{fig:concurrent-fib} presents the results of different thread pool division strategies in comparison to an \enquote{Optimal} allocation.
In Figure~\ref{fig:concurrent-fib-latency} we observe that the collaborative friendly pool does improve latency of the program compared to a simple static thread pool.
However, Figure~\ref{fig:concurrent-fib-throughput} highlights that the throughput of the friendly pool does not match that of the ignorant pool.
This drop in throughput is due to the overhead of the \enquote{frontend} thread generating the work, which is masked when more worker threads are active, as they enable concurrent processing of the requests.
This decreased throughput is also present for the optimal allocation of threads (one per CPU).

To traverse the trade-off space between latency and throughput we can augment the formula for the number of active threads by a variable \(O\).
This is an \emph{overcommitment factor}, specifying by what amount the number of active threads should overcommit the CPUs and is intended to be tuned for different applications.
Figure~\ref{fig:concurrent-overcommit} shows the impact of choosing some sample \emph{overcommitment factor} values, highlighting the trade-off between tail latency and throughput and the space that this parameter enables the user to traverse.

We believe that this approach of making the thread pool dynamic is scalable to current language runtimes for lightweight threading.
They follow the typical pattern of having worker threads that are CPU-bound and blocking threads that consume negligible CPU, waiting for operations to complete.
The lightweight threads can then be multiplexed over the currently active worker threads, enabling improved latency.

%% file: related.tex
\section{Related Work}

Janssen~\cite{adaptivethreadpool} provides a motivation for moving towards dynamic threadpools that take into account multiple dimensions in order to maximise performance, particularly in the presence of neighbours.
However, the performance in focus is throughput, not tail latency as we argue for.
Additionally, our work focuses purely on tuning the threadpool for CPU usage.

Grand central dispatch~\cite{grandcentraldispatch} for Darwin platforms acts as a per-application threadpool manager.
It presents interfaces such as queues for the application to schedule work in multiple classes, from which the dispatch runtime can order execution on OS threads.
The runtime monitors the CPU configuration to balance work, dynamically scaling the thread pool.
This solution shares similarities in design with our threadpool but is more heavily tied into the language runtime with its work dispatch design.

Huang et al~\cite{adaptburstable} make similar arguments to us focusing on burstable containers, they however use CPU sets to dynamically scale the number of threads.
Instead, we simply park a thread, or send it to sleep, awaiting its next runnable time.
This lets the OS scheduler treat the thread as not runnable, rather than requiring more user-level interventions.

%% file: conclusion.tex
\section{Conclusion}

We have presented an analysis of CPU overcommitment from OS threads and the various ways that this situation can arise, focusing on CPU quotas and ignorance of neighbour CPU usage.
We presented a new threadpool design, the \emph{friendlypool}, that dynamically scales the number of active worker threads to match the number of available CPUs and showed that it significantly reduced latencies, by up to \(6.7\times\). %
However, it comes at the cost of some throughput, up to \(1.4\times\), which we mitigate through an overcommitment factor, giving the user an option for traversing the trade-off of latency and throughput. %
This threadpool design is applicable to modern lightweight thread runtimes, enabling the potential for reducing tail latencies from the core of applications in a scalable manner.